\documentclass[twocolumn,letterpaper,pre,showpacs,superscriptaddress]{revtex4}
\usepackage{amsmath}
\usepackage{graphicx}
\usepackage{amssymb}
\usepackage{hyperref}
\usepackage{verbatim}

\newcommand{\Hsh}{H_\mathrm{sh}}
\newcommand{\eref}[1]{Eq.~(\ref{#1})}
\newcommand{\rref}[1]{(\ref{#1})}
\newcommand{\be}{\begin{equation}}
\newcommand{\ee}{\end{equation}}

\begin{document}

\title{Superheating field of superconductors within Ginzburg-Landau theory}

\author{Mark K. Transtrum}
\affiliation{Laboratory of Atomic and Solid State Physics, Cornell University, Ithaca, New York 14853, USA}

\author{Gianluigi Catelani}
\affiliation{Department of Physics, Yale University, New Haven, Connecticut 06520, USA}

\author{James P. Sethna}
\affiliation{Laboratory of Atomic and Solid State Physics, Cornell University, Ithaca, New York 14853, USA}

\begin{abstract}
We study the superheating field of a bulk superconductor within
Ginzburg-Landau theory, which is valid near the critical temperature.
We calculate, as functions of the Ginzburg-Landau parameter $\kappa$,
the superheating field $\Hsh$ and the critical momentum
$k_c$ characterizing the wavelength of the instability of the Meissner state to flux penetration.
By mapping the two-dimensional
linear stability theory into a one-dimensional eigenfunction problem
for an ordinary differential equation, we solve the problem numerically.
We demonstrate agreement between the numerics and analytics, and show convergence
to the known results at both small and large $\kappa$. We discuss the implications
of the results for superconducting RF cavities used in particle accelerators.
\end{abstract}


\maketitle

\section{Introduction}

One of the primary features of superconductivity is the Meissner effect
--- the expulsion of a weak magnetic field from a bulk superconducting
material \cite{Tinkham2004}. For sufficiently large magnetic fields, the Meissner state
becomes unstable, and the system undergoes a phase transition. The
exact nature of the transition depends on the so-called Ginzburg-Landau
parameter, $\kappa=\lambda/\xi$, where $\lambda$ is the London penetration
depth and $\xi$ the superconducting coherence length. Type I superconductors,
characterized by small $\kappa$, transition from the Meissner
state into a normal metal state for magnetic fields above the thermodynamic
critical field, $H_{c}$. Type II superconductors, with larger $\kappa$,
instead transition into a superconducting state with vortices
above the first critical field $H_{c1}$. This state is stable
up to a second critical field $H_{c2}$, above which the metal becomes
normal. For any superconductor,
however, the Meissner
superconducting state is metastable, persisting up to the superheating field $\Hsh$,
well above $H_{c}$ or $H_{c1}$ (for type I and II superconductors,
respectively).
The main goal of the present work is the calculation of $\Hsh$ as function
of $\kappa$ for superconductors near the critical temperature $T_c$ where Ginzburg-Landau
theory is applicable (we remind that within Ginzburg-Landau theory,
the transition from type I to type II superconductors is at $\kappa = 1/\sqrt{2}$).

The metastability of the Meissner state is of interest in the design
of resonance RF cavities in particle accelerators, where $\Hsh$ places
a fundamental limit on the maximum accelerating field \cite{Padamsee}. As type II superconducting
materials are being considered in cavity designs,
a precise calculation of $\Hsh$ in this regime is of value. One
must note, however, that operating temperatures of superconducting
RF cavities are well below the critical temperature $T_{c}$ and that at these low temperatures
Ginzburg-Landau theory is
not quantitatively valid. The numerical techniques developed here
are also being used within the Eilenberger formalism to address these
lower temperatures \cite{inprep}. Using this formalism, the limit $\kappa \to \infty$  was
studied in Ref.~\cite{Catelani2008} for arbitrary temperature.

Much work has already been done in calculating the superheating field
within Ginzburg-Landau theory \cite{Gennes1965,Galaiko1966,Kramer1968,Kramer1973,
Fink,Christiansen,Chapman1995,Dolgert1996}.
The problem is formulated as follows: the superconductor occupies
a half space with a magnetic field applied parallel to the surface.
The order parameter
and vector potential are functions of the distance from the surface and can
be found by solving a boundary value problem of ordinary
differential equations. The superheating field is then the largest
magnetic field for which the corresponding solution is a local minimum
of the free energy. For small values of $\kappa$, the
superheating field corresponds to the largest magnetic field for which
a nontrivial solution to the Ginzburg-Landau equations exist \cite{Dolgert1996},
as the instability does not break translational invariance. However,
as $\kappa$ increases the one-dimensional solution is unstable
to two-dimensional perturbations, resulting in a lower estimate of
$\Hsh$ as first shown in Ref.~\cite{Galaiko1966}. The task at
hand is to find which perturbations destroy the Meissner state and
at which value of the applied magnetic field they first become unstable.

The calculation of $\Hsh$ is therefore a linear stability analysis of the
coupled system of superconducting order parameter and
vector potential.
For a given configuration, we study its stability to arbitrary two-dimensional perturbations
by considering the second variation of the free energy: if the second
variation is positive definite for all possible perturbations then
the solution is (meta)stable.
The second variation can be expressed as
a Hermitian operator acting on the perturbations,
so it is sufficient to show that the eigenvalues of this
operator are all positive.
By expanding the perturbations in Fourier
modes parallel to the surface, the eigenvalue problem can once again
be translated into a boundary value problem of an ordinary differential
equation. The eigenvalues now depend upon the wave-number of the Fourier
mode, but can otherwise be solved in the same way as the Ginzburg-Landau
equations. The superheating field is then the largest applied
magnetic field for which the smallest eigenvalue is positive for all
Fourier modes.

The present stability analysis is more challenging than many such
calculations, as the instability destabilizes an interface
with a pre-existing depth-dependence of field and superconducting
order parameter. As described above,
we map the partial differential equation for the unstable
mode into an eigenvalue analysis for a family of one-dimensional ordinary
differential equations (as originally suggested, but not implemented, in Ref.~\cite{Kramer1968}).
This technique could be useful in a variety of other linear stability
calculations \cite{Brower,Boden,Ben-Jacob}, replacing thin interface approximations
with a microscopic depth-dependent treatment of the destabilizing interface.

The paper is organized as follows:
in the next section we present the Ginzburg-Landau free energy and
the differential equations to be studied for the stability analysis.
In Sec.~\ref{sec:Numerical-Results}
we give some details about the numerical calculations and our main results. In Sec.~\ref{sec:Conclusions}
we discuss the implications of the results for accelerator cavity
design and outline future research directions. In Appendices
we derive analytic formulas, valid at large $\kappa$, which we compare against the numerics.

\section{Ginzburg-Landau theory and stability analysis}
\label{sec:Ginzburg-Landau-theory}

The Ginzburg-Landau free energy for a superconductor occupying the
half space $x>0$ in terms of the magnitude of the superconducting
order parameter $f$ and the gauge-invariant vector potential $\mathbf{q}$
is given by
\begin{eqnarray}
\mathcal{F}[f,\mathbf{q}] & = & \int_{x>0}d^{3}r\Big\{\xi^{2}(\nabla f)^{2}+\frac{1}{2}(1-f^{2})^{2}
\nonumber \\
 &  & +f^{2}\mathbf{q}^{2}+(\mathbf{H}_{a}-\lambda\nabla\times\mathbf{q})^{2}\Big\},
\end{eqnarray}
where $\mathbf{H}_{a}$ is the applied magnetic field (in units of
$\sqrt{2}H_{c}$), $\xi$ is the Ginzburg-Landau coherence length, and
$\lambda$ is the penetration depth. Note that after choosing the unit of
length, the only remaining free parameter in the theory is the ratio
of these two characteristic length scales, the Ginzburg-Landau parameter
$\kappa=\lambda/\xi$. The magnetic field inside of the superconductor
is given by $\mathbf{H} =\lambda\nabla\times\mathbf{q}$.

We take the applied field to be oriented along the $z$-axis
$\mathbf{H}_{a} = (0,0,H_a)$, and the order parameter $f=f(x)$ to depend only on the distance from
the superconductor's surface. We have assumed that the order parameter is real
and further parametrize the vector potential as $\mathbf{q}=(0,q(x),0)$,
which fixes the gauge. The Ginzburg-Landau equations that extremize
$\mathcal{F}$ with respect to $f$ and $\mathbf{q}$ are
\begin{equation}\begin{split}
&\xi^{2}f''-q^{2}f+f-f^{3}=0,\\
&\lambda^{2}q''-f^{2}q=0,
\end{split}\label{eq:1DGL}
\end{equation}
and with our choices
$
H=\lambda q'
$. Hereafter we use primes to denote derivatives with respect to $x$.

The boundary conditions at the surface derive from the requirement
that the magnetic field be continuous, $q'(0)=H_{a}/\lambda$, and
that no current passes through the boundary, $f'(0)=0$. We also require
that infinitely far from the surface the sample is completely superconducting
with no magnetic field, giving us $f(x)\rightarrow1$ and $q(x)\rightarrow0$
as $x\rightarrow\infty$.
In the limits $\kappa \to 0$ and $\kappa \to \infty$, Eqs.~(\ref{eq:1DGL}) can be explicitly solved
perturbatively, see Ref.~\cite{Dolgert1996} and Appendix~\ref{app:op}, respectively. For
arbitrary $\kappa$ they can be solved numerically
via a relaxation method, as we discuss in Sec.~\ref{sec:Numerical-Results}.

For a given solution $(f,\mathbf{q})$ we consider the second variation
of $\mathcal{F}$ associated with small perturbations $f\rightarrow f+\delta f$
and $\mathbf{q}\rightarrow\mathbf{q}+\delta\mathbf{q}$ given by
\begin{eqnarray}
\delta^{2}\mathcal{F} & = & \int_{x>0}d^{3}r\Big\{\xi^{2}(\nabla\delta f)^{2}+
4f\delta f\mathbf{q}\cdot\delta\mathbf{q}+f^{2}\delta\mathbf{q}^{2}\nonumber \\
&  & (3f^{2}+\mathbf{q}^{2}-1)\delta f^{2}+\lambda^{2}(\nabla\times\delta\mathbf{q})^{2}\Big\}.
\label{eq:d2F}
\end{eqnarray}
If the expression in \eref{eq:d2F} is positive for all possible
perturbations, then the solution is stable. Since our solution $(f,\delta\mathbf{q})$
depends only on the distance from the boundary (and is therefore translationally
invariant along the $y$ and $z$ directions), we can expand the
perturbation in Fourier modes parallel to the surface. As shown
in Ref.~\cite{Kramer1968}, we can restrict our attention
to perturbations independent of $z$ and write
\be\begin{split}
\delta f (x,y) & =\delta\tilde{f}(x)\cos ky, \\
\delta\mathbf{q}(x,y) & =(\delta\tilde{q}_{x}\sin ky,\delta\tilde{q}_{y}\cos ky,0), \label{eq:fouriermodes}
\end{split}\ee
where $k$ is the wave-number of the Fourier mode. The remaining Fourier components (corresponding to replacing $\cos \to \sin$ and vice-versa in Eq.~\ref{eq:fouriermodes}) are redundent as they decouple from those given in Eq.~\ref{eq:fouriermodes} and satisfy the same differential equations derived below.

After substituting
into the expression (\ref{eq:d2F}) for the second variation and integrating by parts,
we arrive at\begin{widetext}
\begin{equation}
\delta^{2}\mathcal{F}=\int_{0}^{\infty}dx\left(\begin{array}{ccc}
\delta\tilde{f} & \delta\tilde{q}_{y} & \delta\tilde{q}_{x}\end{array}\right)\left(\begin{array}{ccc}
-\xi^{2}\frac{d^{2}}{dx^{2}}+q^{2}+3f^{2}+\xi^{2}k^{2}-1 & 2fq & 0\\
2fq & -\lambda^{2}\frac{d^{2}}{dx^{2}}+f^{2} & -\lambda^{2}k\frac{d}{dx}\\
0 & \lambda^{2}k\frac{d}{dx} & f^{2}+\lambda^{2}k^{2}\end{array}\right)\left(\begin{array}{c}
\delta\tilde{f}\\
\delta\tilde{q}_{y}\\
\delta\tilde{q}_{x}\end{array}\right).\label{eq:d2FOperator}
\end{equation}
\end{widetext}

The matrix operator in \eref{eq:d2FOperator} is self-adjoint,
and the second variation will be positive definite if its eigenvalues
are all positive.
In the eigenvalue equations for this operator, the function $\delta\tilde{q}_{x}$ can be solved for
algebraically. The resulting differential equations for $\delta\tilde{f}$
and $\delta\tilde{q}_{y}$ are
\begin{eqnarray}
-\xi^{2}\delta\tilde{f}''+(3f^{2}+q^{2}-1+\xi^{2}k^{2})\delta\tilde{f}+2fq\delta\tilde{q}_{y}
&=& E\delta\tilde{f}, \nonumber \\ & &
\label{eq:eigen1}\end{eqnarray}
and
\be
-\lambda^{2}\frac{d}{dx}\left[\frac{f^{2}-E}{f^{2}+\lambda^{2}k^{2}-E}\delta\tilde{q}_{y}'\right]
+f^{2}\delta\tilde{q}_{y}+2fq\delta\tilde{f}  =  E\delta\tilde{q}_{y},
\label{eq:eigen2}\ee
where $E$ is the stability eigenvalue. Note that by decomposing in Fourier
modes, we have transformed the two-dimensional problem into a one-dimensional
eigenvalue problem. Numerically, it can be solved by the same relaxation method
as the Ginzburg-Landau equations -- see Sec.~\ref{sec:Numerical-Results}.
The boundary conditions associated
with the eigenvalue equations derive from the same physical requirements
previously discussed: we require $\delta\tilde{f}'(0)=0$,
since no current may flow through the boundary, and $\delta\tilde{q}_{y}'(0)=0$,
since the magnetic field must remain continuous. Additionally, we
require $\delta\tilde{f}(x)\rightarrow0$ and $\delta\tilde{q}_{y}(x)\rightarrow0$
as $x\rightarrow\infty$. There is also an arbitrary overall normalization,
which we fix by requiring $\delta\tilde{f}(0)=1$.

The stability eigenvalue will depend on the solution of the Ginzburg-Landau
equations, i.e., the applied magnetic field $H_{a}$, and the Fourier
mode $k$ under consideration. The problem at hand is to find the
applied magnetic field and Fourier mode for which the smallest eigenvalue
first becomes negative, which is the case if the following two conditions
hold:
\be\label{eq:eigen_cond}
E=0 \, , \qquad \quad \frac{dE}{dk}=0.
\ee
The value
of the magnetic field at which these conditions are met is the superheating
field $\Hsh$, and the corresponding wave-number is known as the critical momentum $k_{c}$.
In the next section
we discuss in more detail the
numerical approach used to calculate these two quantities.

\section{Numerical Results\label{sec:Numerical-Results}}

As explained in the previous section, the calculation of the superheating field
comprises two main steps: (1) solving the Ginzburg-Landau equations \rref{eq:1DGL} and
(2) solving the eigenvalue problem \rref{eq:eigen1}-\rref{eq:eigen2} with conditions \rref{eq:eigen_cond}.
To solve these equations we employ a relaxation method. The basic
scheme is to replace the ordinary differential equations with a set
of finite difference equations on a grid. From an initial guess to
the solution, the method iterates using Newton's method to relax to
the true solution \cite{Press2007}. The grid is chosen with a high density of points
near the boundary, with the density diminishing approximately as the
inverse distance from the boundary. This is similar to the scheme
used by Dolgert \textit{et al.} \cite{Dolgert1996} to solve the Ginzburg-Landau
equations for type I superconductors.

For $\kappa$ near the type I/II transition, the relaxation method typically converges without much
difficulty.  In the limiting cases that $\kappa$ becomes either very large or small, however, the grid
spacing must be chosen with care to achieve convergence.  The eigenfunction equations are particularly
sensitive to the grid choice. This is not surprising, since in either limit there are two well-separated
length scales. For example, using units $\lambda = 1$ and $\xi = 1/\kappa$, we find that a grid with density
\be
\rho(x) = \frac{150 \kappa}{1 + 25\kappa \\ x} \label{eq:rho}
\ee
leads to convergence for $\kappa$ as high as 250.  The grid points are then be generated recursively $x_{i+1} = x_i + 1/\rho(x_i)$ with $x_0 = 0$.  We find that if the grid is not sufficiently sparse at large $x$, the relaxation method fails, presumably due to rouding errors.  On the other hand if it is too sparse, the finite difference equations poorly approximate the true differential equation.  Fortunately, the method converges quickly, allowing us to explore the density by trial and error, as we have done to get Eq.~\ref{eq:rho}.

\begin{figure}[t]
\includegraphics[scale=0.33]{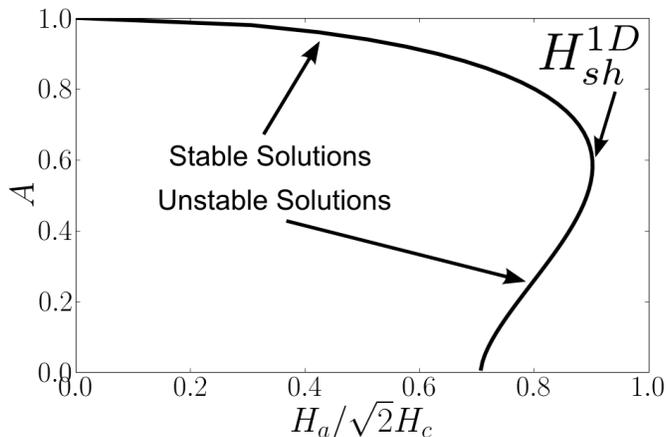}
\caption{\label{fig:HvsA}\textbf{Solving for the 1-D Ginzburg-Landau superheating field}. By fixing value of the order parameter at the surface,
we implicitly define the applied magnetic field $H_{a}(A)$ at the
surface. This definition produces a branch of unstable solutions,
but guarantees that our equations will have a solution for all guesses
of $A$. The {}``nose'' of the curve occurs at the $H_{\mathrm{sh}}^{1D}$,
the superheating field ignoring two-dimensional fluctuations, and
is the largest $H_{a}$ for which a nontrivial solution to \eref{eq:1DGL}
can be found. This example was calculated for $\kappa=1$, and $H_{\mathrm{sh}}^{1D}\approx0.9$.}
\end{figure}

In solving Eq.~\ref{eq:1DGL}, if a sufficiently large value for the applied magnetic
field is used, there may not be a nonzero solution to the Ginzburg-Landau
equations and the relaxation method will often fail to converge, indicating
that the proposed $H_{a}$ is above the actual superheating field.
In practice, therefore, it is more convenient to replace the boundary
condition $q'(0)=\lambda H_{a}$ with a condition on the value of
the order parameter $f(0)=A$, which then implicitly defines the applied
magnetic field as a function of $A$, $H_{a}(A)$. This has the advantage
that $H_{a}(A)$ is a differentiable function of $A$, as is the stability
eigenvalue, improving the speed
and accuracy of the search for the superheating field. The drawback
to this approach is that $H_{a}(A)$ is not single-valued, with an unstable
branch of solutions as illustrated in Fig.~\ref{fig:HvsA}.
For the problem at hand, this turns out to be straightforward to address
since we determine the stability of each solution in the second step.
To achieve the conditions in \eref{eq:eigen_cond}, we vary both the Fourier mode, $k$,
and the value of the order parameter at the surface $A$.

\begin{figure}
\includegraphics[scale=0.33]{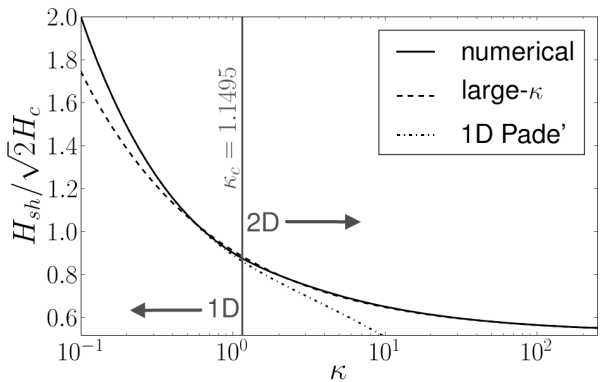}\caption{\label{fig:Hsh}Numerically calculated $\Hsh$
and corresponding analytical approximations [Eqs.~\rref{eq:pade} and \rref{eq:Hshapp}] versus
the Ginzburg-Landau parameter $\kappa$.}
\end{figure}

\begin{figure}
\includegraphics[scale=0.33]{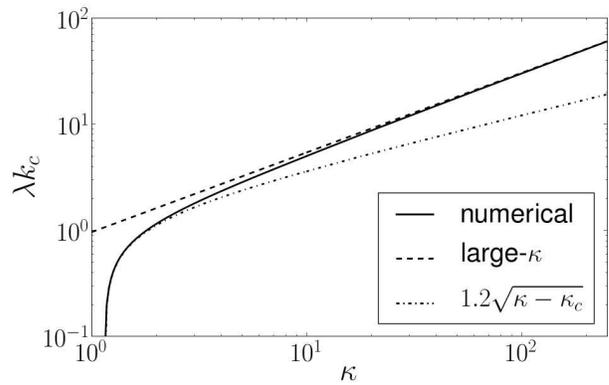}\caption{\label{fig:kccomparison}
Comparison between the numerical and asymptotic critical momentum $k_{c}$, Eq.~(\ref{eq:kc_asym}).
The approximate behavior of $k_c$ near $\kappa_c$ (dot-dashed) is given in \eref{eq:kckac}.}
\end{figure}

\begin{figure}[!t]
\includegraphics[scale=0.33]{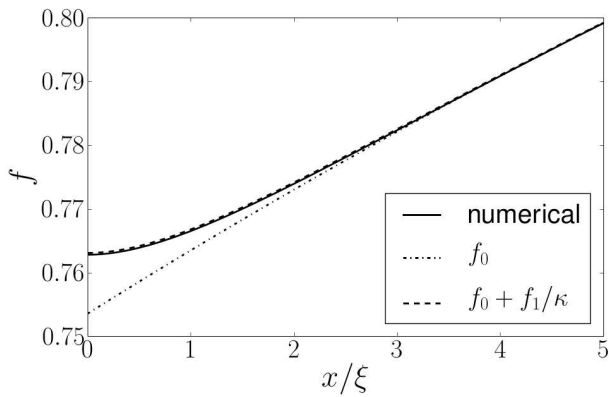}\caption{\label{fig:fprofile}Profile of
the order parameter at $\Hsh$ for
$\kappa=50$ together with analytic approximations given in Eqs.~\rref{eq:Q0F0} and (\ref{eq:QFapp}).}
\end{figure}

\begin{figure}
\includegraphics[scale=0.33]{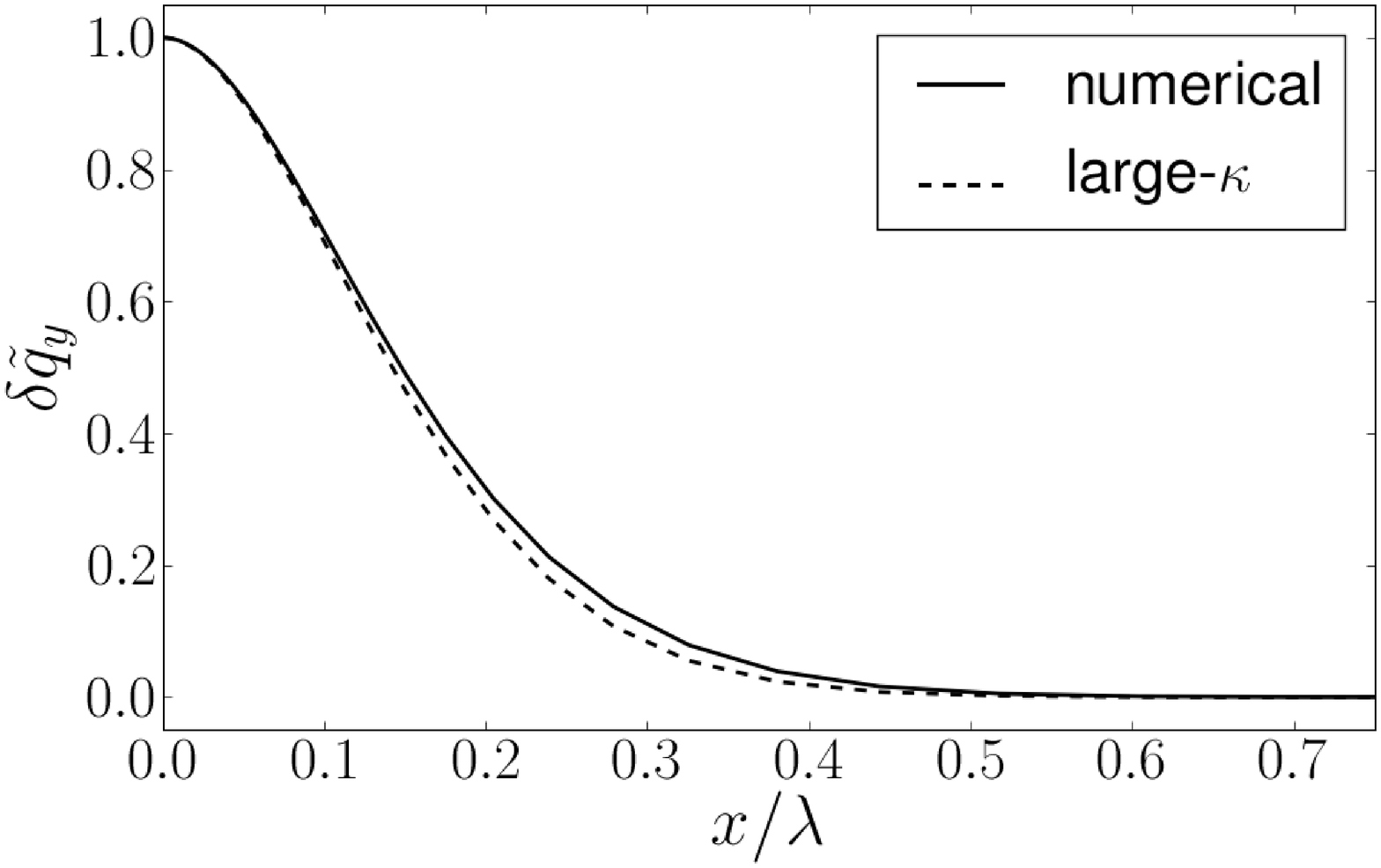}
\caption{\label{fig:dqprofile}Numerical profile (solid line) of the critical perturbation $\delta\tilde{q}_y$
determining $\Hsh$ for $\kappa=50$ compared to the large-$\kappa$ perturbative
result (dashed), \eref{eq:dqypert}.
}
\end{figure}

The results of the procedure described above are summarized
in Figs.~\ref{fig:Hsh}-\ref{fig:dqprofile}, where we also compare them with analytical estimates
which, for large values of $\kappa$, are derived in Appendices.
In Fig.~\ref{fig:Hsh} we plot the numerically calculated superheating field as a function of
$\kappa$ (solid line). The vertical line at $\kappa_c \simeq 1.1495$ separates the regimes
of one-dimensional (1D, $k=0$) and two-dimensional (2D, $k\neq 0$) critical perturbations. We have checked
the value of $\kappa_c$ both by assuming 2D perturbations and finding when their critical momentum goes to
zero and by assuming 1D perturbation
and finding when the coefficient of the term quadratic in momentum in the second variation of the free
energy vanishes;
these methods lead to the same value within our numerical accuracy of $10^{-4}$. Our value of $\kappa_c$ is
higher than previous estimates, which ranged from 0.5 \cite{Kramer1968}
to 1.10 \cite{Fink} and 1.13($\pm$0.05) \cite{Christiansen}.  $\kappa_c$ is larger than the boundary $\kappa=1/\sqrt{2}$ separating type I from type II superconductors.  Type II superconductors for which $\kappa < \kappa_c$ become unstable via a spatially uniform invasion of magnetic flux.  Additionally, we find that superheated type II superconductors with $\kappa < 0.9192$ can transition directly into the normal state since the corresponding $\Hsh$ is larger than the second critical field $H_{c2}$.  

For $\kappa<\kappa_c$ the instability is due to 1D
perturbations. In this regime, the Pad\'e approximant
\be\label{eq:pade}
\frac{\Hsh(\kappa)}{\sqrt{2}H_c} \approx 2^{-3/4}\kappa^{-1/2}
\frac{1+4.6825120\kappa+3.3478315\kappa^2}{1+4.0195994\kappa+1.0005712\kappa^2}
\ee
derived in
Ref.~\cite{Dolgert1996} (dot-dashed line) gives a good approximation to the actual $\Hsh$, with deviation
of less than about 1.5~\%. In the opposite case $\kappa > \kappa_c$, 2D perturbations are
the cause of instability and the superheating field is approximately given by \cite{Christiansen} (dashed line)
\be\label{eq:Hshapp}
\frac{\Hsh(\kappa)}{\sqrt{2}H_c} \approx \frac{\sqrt{10}}{6}+\frac{0.3852}{\sqrt{\kappa}}.
\ee
Equation~(\ref{eq:Hshapp}), derived in Appendix~\ref{app:Analytic-Approx},
is also a good approximation, deviating at most about 1~\% from the numerics.
Therefore, our numerics show that the simple analytical formulas for $\Hsh$
in Eqs.~\rref{eq:pade} and \rref{eq:Hshapp} can be used to accurately estimate the superheating field for
arbitrary value of the Ginzburg-Landau parameter, when used in their respective validity regions.

In Fig.~\ref{fig:kccomparison} we show the numerical result for the critical momentum $k_c$ versus
$\kappa$. We see that $k_c \to 0$ as $\kappa \to \kappa_c$ from above. Near $\kappa_c$, the behavior
of $k_c$ is reminiscent of that of an order parameter near a second-order phase transition:
\be\label{eq:kckac}
k_c \simeq 1.2\sqrt{\kappa - \kappa_c},
\ee
where the prefactor has been estimated by fitting the numerics.
The dashed line is the
asymptotic formula \cite{Christiansen} (see also Appendix~\ref{app:Analytic-Approx})
\be\label{eq:kc_asym}
\lambda k_c\approx 0.9558 \kappa^{3/4}
\ee
which captures correctly the large-$\kappa$ behavior.

In Fig.~\ref{fig:fprofile}
we present a typical solution for the order parameter near the surface at $H=\Hsh$ for a
large value of $\kappa$, along with the analytic approximations presented
in Appendix~\ref{app:op}. The zeroth order approximation $f_0$, \eref{eq:Q0F0},
fails near the surface, as it does not satisfy the boundary condition
$f'(0)=0$. On the other hand, including the first order correction
in $1/\kappa$, \eref{eq:QFapp}, leads to excellent agreement with the numerics.
Finally, in Fig.~\ref{fig:dqprofile}
we show a typical example of the depth dependence of the perturbation $\delta\tilde{q}_y$
at the critical point where the solution first becomes unstable. We find again good agreement
between numerical and perturbative calculations.

It is interesting to compare the wavelength of the critical perturbation, $2\pi/k_c$,
with the Abrikosov spacing $a$ for the arrangement of vortices at
the superheating field: the naive expectation is that the initial flux penetrations
represent nuclei for the final vortices. Kramer argues that this
picture is incorrect since the initial flux penetrations do
not have supercurrent singularities and do not carry a fluxoid
quantum \cite{Kramer1968}. 

\begin{figure}
\includegraphics[scale=0.33]{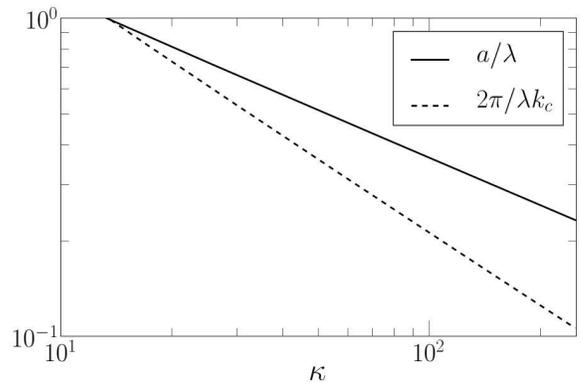}

\caption{\label{fig:Abrikosov}The wavelength of the critical perturbation ($2\pi/k_{c}$) and the
Abrikosov vortex spacing ($a$) calculated at the superheating field both vanish at large $\kappa$, although the
former diminishes much more quickly.}

\end{figure}

We find the numerical discrepancies
between the two lengths further support Kramer's argument.
In the weakly type-II regime ($\kappa \sim 1$), the initial flux
penetration is from infinitely long wavelengths ($k_c = 0$);  in
contrast, the final vortex state has a very high density, since
$\Hsh \sim H_{c2}$ \cite{note1}. In the strongly type-II limit ($\kappa
\rightarrow \infty$) both the inverse momentum and
the Abrikosov spacing (evaluated at the superheating field) vanish, but at
different rates, with $1/k_c \sim \kappa^{-3/4}$ while the Abrikosov
spacing $a \sim \kappa^{-1/2}$ at $\Hsh$ \cite{note2}, see
Fig.~\ref{fig:Abrikosov}. These results suggest that there is no
immediate connection between the initial penetration and the
final vortex array. A dynamical simulation could explore the
transition between the initial penetration and the final vortex
state, similar to that done by Frahm \textit{et.~al} for the transition from
the normal state to the vortex state \cite{Frahm1991}.

\section{Summary and outlook\label{sec:Conclusions}}

In this paper, we have numerically calculated within Ginzburg-Landau theory the superheating field
$\Hsh$ of superconductors by mapping the linear stability threshold
onto an eigenfunction problem, and we have
shown that analytic approximations are in good agreement with the numerical
results. The technique of mapping the linear stability problem onto
a one-dimensional eigenfunction problem is potentially a useful technique,
and we hope others find useful applications of the methods
described here.

One of the primary motivations for this work is the application to
RF cavities in particle accelerators, where the maximum accelerating
field is limited by $\Hsh$. While the results presented here provide
good estimates of $\Hsh$ for many materials of interest near the critical temperature $T_c$,
we emphasize that the operating temperature of these cavities are typically well below $T_{c}$,
where Ginzburg-Landau theory is not quantitatively accurate. The techniques presented here
can be applied to Eilenberger theory to more accurately determine
$\Hsh$ at low temperatures. The Eilenberger approach
has already been used \cite{Catelani2008} to evaluate $\Hsh(T)$ at any temperature in the infinite $\kappa$ limit
for clean superconductors, and work is in progress to address low temperatures for
finite $\kappa$ \cite{inprep}.

\acknowledgments

The authors would like to thank Hasan Padamsee, Georg Hoffstaetter,
and Matthias Liepe for helpful discussion. This work was supported
by NSF grant number DMR-0705167 (MKT \& JPS), by the Department of Energy under contract de-sc0002329
(MKT), and by Yale University (GC).

\appendix

\section{Order parameter and vector potential in the large $\kappa$ limit}
\label{app:op}

In this Appendix, we derive solutions to the Ginzburg-Landau
equations \eref{eq:1DGL} valid in the large-$\kappa$
limit. For convenience, we work in units $\lambda=1$, $\xi=1/\kappa$.
As a first step, we consider the limit
$\kappa\to\infty$. Then Eqs.~\ref{eq:1DGL} reduce to
\begin{equation}
\begin{split}
q_{0}'' & =f_{0}^{2}q_{0}, \\
0 & =f_{0}\left(f_{0}^{2}-1+q_{0}^{2}\right),
\end{split}
\end{equation}
with solution \cite{Gennes1965}
\begin{equation}
\begin{split}
q_{0}(x)=-\frac{\sqrt{2}}{\cosh(x+\ell)}, \\
f_{0}(x)=\sqrt{1-q_{0}^{2}(x)},
\end{split}
\label{eq:Q0F0}\end{equation}
where the parameter $\ell$ is determined by the field at the surface via
\begin{equation}
H_{a}=q_{0}'(0)=\frac{\sqrt{2}\sinh\ell}{\cosh^{2}\ell}.
\label{eq:H0l}
\end{equation}

The above solution satisfies the boundary conditions at infinity,
but it cannot satisfy the boundary condition for $f$ at the surface.
An approximate solution, valid at finite but large $\kappa$, which
satisfies all boundary conditions can be obtained by boundary layer
theory. We follow the approach of Ref.~\cite{Chapman1995}, so we
only sketch the steps of the calculation. Note that away from the
thermodynamic critical field, the scaling is different than that used
in Ref.~\cite{Chapman1995}: there the expansion is in powers of
$\kappa^{-\alpha}$ and the inner variable is $X=\kappa^{\alpha}x$
with $\alpha=2/3$, here we use $\alpha=1$: 
\begin{equation}
\begin{split}q & =q_{0}+\frac{1}{\kappa}q_{1}+\ldots\\
f & =f_{0}+\frac{1}{\kappa}f_{1}+\ldots\end{split}
\end{equation}
Substituting into Eqs.~(\ref{eq:1DGL}), we find the following ``outer
layer'' equations for $q_{1}$ and $f_{1}$:
\begin{equation}
\begin{split}q_{1}'' & =2f_{0}f_{1}q_{0}+f_{0}^{2}q_{1} ,\\
0 & =f_{1}(3f_{0}^{2}-1+q_{0}^{2})+2f_{0}q_{0}q_{1}\end{split}
\end{equation}
which have the simple solutions $f_{1}=q_{1}=0$. For the inner layer,
we introduce the variable $X=\kappa x$ and find the equations
\begin{equation}
\begin{split}\tilde{f}_{0}'' & =\tilde{f}_{0}\left(\tilde{f}_{0}^{2}-1+\tilde{q}_{0}^{2}\right),\\
\tilde{q}_{0}'' & =0,\end{split}
\label{eq:ieq0}\end{equation}
and
\begin{equation}
\begin{split}\tilde{f}_{1}'' & =\tilde{f}_{1}\left(3\tilde{f}_{0}^{2}-1+\tilde{q}_{0}^{2}\right)
+2\tilde{f}_{0}\tilde{q}_{0}\tilde{q}_{1},\\
\tilde{q}_{1}'' & =0,\end{split}
\label{eq:ieq1}\end{equation}
where we use tildes to denote functions of the inner variable $X$.
Equations \ref{eq:ieq0} have constant solutions
\begin{equation}
\tilde{q}_{0}=-b\,,\quad\tilde{f}_{0}=\sqrt{1-b^{2}},
\label{eq:tQ0}
\end{equation}
while from the second of Eqs.~(\ref{eq:ieq1}) and the boundary
conditions we get
\begin{equation}
\tilde{q}_{1}=H_{a}X \, .
\end{equation}
Then the first of Eqs.~(\ref{eq:ieq1}) becomes
\begin{equation}
\tilde{f}_{1}''=2(1-b^{2})\tilde{f}_{1}-2b\sqrt{1-b^{2}}H_{a}X,
\end{equation}
with solution
\begin{equation}
\tilde{f}_{1}=\frac{bH_{a}}{\sqrt{1-b^{2}}}X+Ae^{-\sqrt{2}\sqrt{1-b^{2}}X}+Be^{\sqrt{2}\sqrt{1-b^{2}}X}
\end{equation}
with $A,B$ integration constants. Since $f$ tends to a constant
far from the surface, we set $B=0$. Vanishing of the derivative at
the surface then fixes
\begin{equation}
A=\frac{bH_{a}}{\sqrt{2}(1-b^{2})}
\end{equation}

Next, we match the inner and outer solutions. Comparing Eqs.~(\ref{eq:Q0F0})
and (\ref{eq:tQ0}) we get
\begin{equation}
b=\frac{\sqrt{2}}{\cosh\ell}\label{eq:bl}.
\end{equation}
We can express $b$ in terms of the applied field using \eref{eq:H0l} to find
\be
b = \sqrt{1-\sqrt{1-2H_a^2}}.
\ee
Then, since $f_{1}=q_{1}=0$, we need to compare the linear order
expansion of Eqs.~(\ref{eq:Q0F0}) at small $x$ with $\tilde{q}_{1}/\kappa$
and $\tilde{f}_{1}/\kappa$ at large $X=\kappa x$. Using Eqs.~(\ref{eq:H0l})
and (\ref{eq:bl}), we find that the inner and outer solutions match.
Finally, the uniform approximate solution is \begin{equation}
\begin{split}q(x) & =q_{0}(x),\\
f(x) & =\sqrt{1-q_{0}^{2}(x)}+\frac{1}{\kappa}
\frac{bH_{a}}{\sqrt{2}(1-b^{2})}e^{-\sqrt{2}\sqrt{1-b^{2}}\kappa x}
\end{split}
\label{eq:QFapp}\end{equation}
with corrections of order $1/\kappa^{2}$. In Fig.~\ref{fig:fprofile}
we compare the second of Eq.~(\ref{eq:QFapp}) to numerics.

\section{Superheating field in the large-$\kappa$ limit}
\label{app:Analytic-Approx}

The calculation of the superheating field $\Hsh$ as a function of $\kappa$
for stability with respect to one-dimensional perturbations (i.e., $k=0$) can be found
in Ref.~\cite{Dolgert1996} for $\kappa \to 0$ and Ref.~\cite{Chapman1995} for $\kappa \to \infty$.
The latter calculation, however, is of little physical relevance, as the actual instability
at sufficiently large $\kappa$ is due to two-dimensional perturbations.
Here we present for completeness (albeit in a different form)
Christiansen's perturbative calculation \cite{Christiansen} of the true superheating field
$\Hsh(\kappa)$ for $\kappa\gg 1$.

Our starting point is the following expression for the ``critical'' second variation
of the thermodynamic potential as functional of perturbations $\delta\tilde{f}$, $\delta\tilde{q}_{y}$,
and momentum $k$ [see also Eq.~(10) in Ref.~\cite{Kramer1968}]:
\begin{equation}
\begin{split}\delta^{2}\mathcal{F}= & \int_{0}^{\infty}\!\! dx\Big\{\left[3f^{2}+q^{2}-1+(k/\kappa)^{2}\right]
\delta\tilde{f}^{2}+\kappa^{-2}\delta\tilde{f}'^{2}\\
& \qquad+4fq\delta\tilde{f}\delta\tilde{q}_{y}+f^{2}\delta\tilde{q}_{y}^{2}+(f^{2}+k^{2})^{-1}f^{2}
\delta\tilde{q'}_{y}^{2}\Big\}\end{split}
\end{equation}
It is straightforward to check that variation of this functional with respect to $f$ and $q_{y}$
leads to Eqs.~\rref{eq:eigen1}-\rref{eq:eigen2} with $E=0$ and rescaled units $\lambda=1$.
Kramer estimated that the critical momentum $k\propto\sqrt{\kappa}$.
While we will show that this is not the correct scaling, this form
suggests to rescale lengths by $1/\sqrt{\kappa}$ by defining $x=w/\sqrt{\kappa}$:
\begin{equation}
\begin{split}\delta^{2}\mathcal{F}= & \int_{0}^{\infty}\!\frac{dw}{\sqrt{\kappa}}\Big\{
\left[3f^{2}+q^{2}-1+(k/\kappa)^{2}\right]\delta\tilde{f}^{2}+\kappa^{-1}\delta\tilde{f}'^{2}\\
& \qquad+4fq\delta\tilde{f}\delta\tilde{q_{y}}+f^{2}\delta\tilde{q}_{y}^{2}
+(f^{2}+k^{2})^{-1}f^{2}\kappa\delta\tilde{q'_{y}}^{2}\Big\}\end{split}
\label{eq:tpr}
\end{equation}
where now prime is derivative with respect to $w$. (Note that although
$k$ has units of inverse length, it is momentum parallel to the surface,
and therefore does not scale with $x$.)

Minimization with respect to $k$ leads to the equation
\begin{equation}
k\int\! dw\left[\frac{\delta\tilde{f}^{2}}{\kappa^{2}}-
\frac{\kappa f^{2}}{(f^{2}+k^{2})^{2}}\delta\tilde{q}_{y}'^{2}\right]=0
\end{equation}
Assuming $k\gg1$, we can neglect $f^{2}\leq1$ in the denominator
and find
\begin{equation}
k^{4}\int\! dw\,\delta\tilde{f}^{2}=\kappa^{3}\int\! dw\, f^{2}\delta\tilde{q}_{y}'^{2}
\end{equation}
which shows that (if our length rescaling is correct) the proper
scaling for the critical momentum is $k\propto\kappa^{3/4}$. If this is true, then
$(k/\kappa)^{2}\propto1/\sqrt{\kappa}$ and $\kappa/k^{2}\propto1/\sqrt{\kappa}$,
which shows that the next to leading order terms in curly brackets
in Eq.~(\ref{eq:tpr}) are proportional to $1/\sqrt{\kappa}$. Therefore,
terms of order $1/\kappa$ can be neglected and, in particular, we can
neglect $\kappa^{-1}\delta\tilde{f}'^{2}$ and use everywhere the
lowest order solution for $f$ and $q$, Eq.~(\ref{eq:Q0F0}). Hence
the approximate functional in the large-$\kappa$ limit is
\begin{equation}
\begin{split}\delta^{2}\mathcal{F}\simeq & \frac{1}{\sqrt{\kappa}}\int_{0}^{\infty}\!\! dw\Big\{
\left[2f_{0}^{2}+(k/\kappa)^{2}\right]\delta\tilde{f}^{2}\\
& \quad+4f_{0}q_{0}\delta\tilde{f}\delta\tilde{q}_{y}+f_{0}^{2}\delta\tilde{q}_{y}^{2}+
k^{-2}f_{0}^{2}\kappa\delta\tilde{q}_{y}'^{2}\Big\}.\end{split}
\label{eq:tplk}\end{equation}

By minimizing \eref{eq:tplk} with respect to $\delta\tilde{f}$,
we find
\begin{equation}
\left[2f_{0}^{2}+(k/\kappa)^{2}\right]\delta\tilde{f}=-2f_{0}q_{0}\delta\tilde{q}_{y},
\end{equation}
and solving for $\delta\tilde{f}$
\begin{equation}\label{eq:tftqy}
\delta\tilde{f}=-\frac{2f_{0}q_{0}\delta\tilde{q}_{y}}{2f_{0}^{2}+(k/\kappa)^{2}}\simeq
-\frac{q_{0}\delta\tilde{q}_{y}}{f_{0}}+
\left(\frac{k}{\kappa}\right)^{2}\frac{q_{0}\delta\tilde{q}_{y}}{2f_{0}^{3}}
\end{equation}
where in the last step we kept only the leading and the next to leading
order terms. Substituting back into Eq.~(\ref{eq:tplk}) gives
\begin{equation}\begin{split}
\delta^{2}\mathcal{F}= & \int_{0}^{\infty}\!\frac{dw}{\sqrt{\kappa}}\bigg[
\left(1-3q_{0}^{2}\right)\delta\tilde{q}_{y}^{2} \\ &+
\left(\frac{k}{\kappa}\right)^{2}\frac{q_{0}^{2}}{f_{0}^{2}}\delta\tilde{q}_{y}^{2}
+\frac{\kappa}{k^{2}}f_{0}^{2}\delta\tilde{q}_{y}'^{2}\bigg]
\end{split}
\label{eq:tpfin}\end{equation}
The first term in square brackets is the leading term. Neglecting
the other terms, since $q_{0}^{2}$ is monotonically decreasing function
of $w$ the variation $q_{y}$ that minimizes the functional is a
$\delta$-function at the surface. Then the condition for the metastability
is
\begin{equation}
1-3q_{0}^{2}(0)=0.
\end{equation}
Using \eref{eq:Q0F0} we obtain
\begin{equation}
\cosh\ell=\sqrt{6}\,,\quad\sinh\ell=\sqrt{5}
\end{equation}
and substituting into \eref{eq:H0l}
\begin{equation}
\Hsh^{\infty}=\frac{\sqrt{10}}{6}
\end{equation}

To calculate the large-$\kappa$ correction, we expand the function $q_0(w)$ in the first term in square
brackets in \eref{eq:tpfin} to linear order,
while $q_0$ and $f_0$ in the subleading terms can be simply evaluated at the surface.
Setting
\be\label{eq:ellkappa}
\ell \simeq \mathrm{arccosh}\sqrt{6} - \frac{c}{\sqrt{\kappa}},
\ee
\be\label{eq:kres}
k = \left(\frac{5}{6}\right)^{1/4}\tilde{k} \kappa^{3/4},
\ee
and using \eref{eq:H0l} we find
\be\begin{split}
\delta^{2}\mathcal{F}=2\sqrt{\frac{5}{6}}\int_{0}^{\infty}\!\frac{dw}{\kappa}\bigg[
\left(-c+ w + \frac{1}{4}\tilde{k}^2\right)\delta\tilde{q}_{y}^{2}
+ \frac{2}{5\tilde{k}^2}\delta\tilde{q}_{y}'^{2} \bigg]
\end{split}\ee
The variational equation for $\delta\tilde{q}_{y}$ derived from this functional has as solution
the Airy function
\be\label{eq:tqysol}
\delta\tilde{q}_{y}(w) = \mathrm{Ai}\left[\left(\frac{5\tilde{k}^2}{2}\right)^{1/3}
\left(w-c+\frac{1}{4}\tilde{k}^2\right)\right]
\ee
Imposing the boundary condition $\delta\tilde{q}_{y}'(0)=0$, we find that for a given $\tilde{k}$ the
lowest possible $c$ is
\be
c = z_0 \left(\frac{5\tilde{k}^2}{2}\right)^{-1/3} + \frac{1}{4}\tilde{k}^2,
\ee
where
\be
z_0 \approx 1.018793 \label{eq:z0}
\ee
is the smallest number satisfying $\mathrm{Ai}'(-z_0) =0$. Finally minimizing $c$ with respect to
$\tilde{k}$ we find
\be\label{eq:tkf}
\tilde{k} = \left(\frac{4}{3}z_0\right)^{3/8}\left(\frac{2}{5}\right)^{1/8}
\ee
and
\be\label{eq:cf}
c =\left(\frac{2}{5}\right)^{1/4}\left(\frac43z_0\right)^{3/4}.
\ee

Substituting \eref{eq:ellkappa} into \eref{eq:H0l} we obtain
\be
\Hsh = \frac{\sqrt{10}}{6} + \frac{2c}{3\sqrt{3\kappa}}
\approx \frac{\sqrt{10}}{6} + \frac{0.3852}{\sqrt{\kappa}}
\ee
and from Eqs.~\rref{eq:kres}, \rref{eq:z0}, and \rref{eq:tkf}
\begin{eqnarray}
k &=& \left( \frac{160}{243} \right)^{1/8} z_0^{3/8} \kappa^{3/4} \nonumber \\
 &\approx& 0.9558 \kappa^{3/4}.
\end{eqnarray}
These results agree with those of Ref.~\cite{Christiansen}.
We compare these two formulas with numerics in Figs.~\ref{fig:Hsh} and
\ref{fig:kccomparison}, respectively.

Finally, fixing the arbitrary normalization of the perturbation by requiring $\delta\tilde{q}_y(0)=0$, using
Eqs.~\rref{eq:tqysol}-\rref{eq:cf}, and restoring dimensions we find
\be\label{eq:dqypert}
\delta\tilde{q}_y(x) = \mathrm{Ai}\left[\left(\frac{10}{3}z_0\right)^{1/4}
\frac{\sqrt{\kappa}x}{\lambda}-z_0\right] \bigg/\mathrm{Ai}[-z_0],
\ee
which shows that the ``penetration depth'' of the perturbation is of the order of
the geometric average of coherence length and magnetic field penetration depth.
This functional form is plotted in Fig.~\ref{fig:dqprofile} for $\kappa =50$ along with the numerically
calculated $\delta\tilde{q}_y$.


\begin{thebibliography}{99}
\bibitem{note2}The vortex spacing $a$ is related to the induction $B$ as $a/\lambda \sim 1/(\kappa B)^{1/2}$,
and at intermediate fields $H$ such that $H_{c1} \ll H \ll H_{c2}$,
$B \approx H$ -- see the reference in the previous footnote. Since
at large $\kappa$ the superheating field tends to a constant, we
find $a/\lambda \sim \kappa^{-1/2}$.
\end{thebibliography}
\end{document}